# Atomic Origin of Spin-Valve Magnetoresistance at the SrRuO$_3$ Grain Boundary

Xujing Li[1,2,‡], Li Yin[3,‡], Zhengxun Lai[3], Mei Wu[1,4], Yu Sheng[5], Lei Zhang[2], Yuanwei Sun[1,4], Shulin Chen[1], Xiaomei Li[2], Jingmin Zhang[1], Yuehui Li[1,4], Kaihui Liu[6,7], Kaiyou Wang[5,8], Dapeng Yu[1,6,9], Xuedong Bai[2,7],*, Wenbo Mi[3],* and Peng Gao[1,4,7],*

[1]Electron Microscopy Laboratory, School of Physics, Peking University, Beijing, 100871, China

[2]Beijing National Laboratory for Condensed Matter Physics and Institute of Physics, Chinese Academy of Sciences, Beijing 100190, China

[3]Tianjin Key Laboratory of Low Dimensional Materials Physics and Preparation Technology, School of Science, Tianjin University, Tianjin 300354, China

[4]International Center for Quantum Materials, Peking University, Beijing, 100871, China

[5]State Key Laboratory of Superlattices and Microstructures, Institute of Semiconductors, Chinese Academy of Sciences, Beijing 100083, China

[6] State Key Laboratory for Artificial Microstructure & Mesoscopic Physics, School of Physics, Peking University, Beijing 100871, China

[7]Collaborative Innovation Centre of Quantum Matter, Beijing 100871, China

[8]College of Materials science and Opto-Electronic Technology, University of Chinese Academy of Sciences, Beijing 100049, P. R. China

[9]Shenzhen Institute for Quantum Science and Engineering (SIQSE), and Department of Physics, Southern University of Science and Technology (SUSTech), Shenzhen 518055, P.R.China.

[‡]These authors contributed equally to this work.
*E-mails: p-gao@pku.edu.cn; miwenbo@tju.edu.cn; xdbai@iphy.ac.cn

# ABSTRACT

Defects ubiquitously exist in crystal materials and usually exhibit a very different nature than the bulk matrix, and hence, their presence can have significant impacts on the properties of devices. Although it is well accepted that the properties of defects are determined by their unique atomic environments, the precise knowledge of such relationships is far from clear for most oxides due to the complexity of defects and difficulties in characterization. Here, we fabricate a 36.8° $SrRuO_3$ grain boundary of which the transport measurements show a spin-valve magnetoresistance. We identify its atomic arrangement, including oxygen, using scanning transmission electron microscopy and spectroscopy. Based on the as-obtained atomic structure, the density functional theory calculations suggest that the spin-valve magnetoresistance is because of the dramatically reduced magnetic moments at the boundary. The ability to manipulate magnetic properties at the nanometer scale via defect control allows new strategies to design magnetic/electronic devices with low-dimensional magnetic order.

**Keywords:** spin-valve, magnetic defects, electron microscopy, grain boundary,

# INTRODUCTION

The altered continuity of atomic bonding at grain boundaries makes the physical properties of these defects significantly different from those of the rest of the bulk matrix. For instance, the grain boundaries of ferromagnetic $Pr_{0.7}Ca_{0.3}MnO_3$ and $La_{2/3}Ca_{1/3}MnO_3$ are paramagnetic due to the possible space charge accumulation causing energy band bending or stress-induced structural disordering[1-3]. The low angle grain boundary of paraelectric $SrTiO_3$ is polarized because of the large strain-gradient-induced flexoelectric effect[4]. In the acceptor-doped yttrium-zirconium oxide ionic conductor, the ionic conductivity of the grain boundary is two orders of magnitude lower than that of the grain owing to the oxygen vacancy depletion layer near the grain boundary[5]. For the solar cell material $CuInSe_2$, the performance of the polycrystalline material is better than that of the single crystal because the electrons at the grain boundaries are not easily recombined with the surrounding holes[6].

The properties of grain boundaries strongly depend on their geometry (e.g., tilt and twist angles between the grains), elemental/charge segregation (e.g., nonstoichiometric ratio, termination surface, and space charge accumulation), and strain conditions (including strain and strain gradient). For example, in superconducting copper oxides, only those grain boundaries with high tilt angles can limit the critical current as a Josephson junction[7,8]. Therefore, determining the atomic structure of grain boundaries and revealing the structure-property relations are vital for grain boundary engineering (via controlling the angle and/or element doping) to improve materials and design devices with novel functions.

In this work, we studied the atomic structure and magnetic properties of SrRuO$_3$ (SRO) grain boundaries by combining advanced scanning transmission electron microscopy (STEM), spectroscopy, density functional theory (DFT) calculations and transport property measurements. As an itinerant ferromagnet, SRO has some intriguing electrical and magnetic properties[9-11]. It has been reported that the grain boundary in SRO can cause substantial negative magnetoresistance, while no tunneling magnetoresistance (TMR) was detected[12], which is very different behavior than that of the grain boundary in ferromagnetic La$_{2/3}$Sr$_{1/3}$MnO$_3$[13]. On the other hand, SRO is widely used in electrodes for the growth of thin films, such as superconductors and ferroelectrics. The microstructure of the grain boundaries in SRO may propagate into the thin films, and thus, the properties of the grain boundaries can certainly significantly influence the interface properties, such as the magnetoelectric coupling. These properties are dictated by the microstructure of the SRO grain boundary. However, the atomic structure of grain boundaries in SRO has rarely been studied, and the properties of grain boundaries and the effects of their presence on thin-film devices are largely unknown.

Here, we fabricate a 36.8° SrRuO$_3$ grain boundary (labeled Σ5(310) [001] SRO GB, where Σ denotes the degree of geometrical coincidence of crystalline interfaces[14], (310) is the grain boundary plane, and [001] is the rotation axis). The transport measurements show spin-valve magnetoresistance at the grain boundary. To reveal the underlying mechanism, we determine the atomic structure (including oxygen positions) by using the recently developed atomically resolved integrated differential phase contrast (iDPC) imaging technique combined with atomically resolved energy dispersive X-ray spectroscopy (EDS) with aberration corrected STEM. Based on the obtained atomic structure, we perform DFT calculations and find that along the grain boundary, the magnetic moments are reduced by ~ 91% on one side and ~ 25% on the other side. The changes in magnetic moments and spin polarization stem from the reconstruction of the Ru $d$ orbital due to the Ru-O octahedron distortion. The substantial reduction of the magnetic moments leads to spin-valve magnetoresistance at the grain boundary. These findings unveil the structure and properties of the grain boundary in a commonly used ferromagnetic electrode SRO, which can help us to understand the effects of such a grain boundary on the magnetic transport properties of SRO and provide new insights into defect engineering for novel magnetic/electric devices.

**RESULTS**

**Design and fabrication of the bicrystal.** The high quality of the SRO boundary was fabricated by growth of an SRO thin film on a SrTiO$_3$ (STO) bicrystal substrate. Fig. 1(a) is a schematic diagram of the SRO film growth on an STO bicrystal substrate. The STO bicrystal with a 36.8° mis-tilted grain boundary was fabricated by thermal diffusion bonding[15], and the SRO thin film was deposited on the STO bicrystal substrate by pulsed laser deposition (PLD)[16]. The experimental details are included in the methods section. The cross-sectional high angle annular dark field (HAADF)-

STEM image in Fig. 1(b) shows that the film thickness is approximately 50 nm. The high-magnification HAADF-STEM image of the sample planar view along the [001] direction in Fig. 1(c) shows the uniform and high quality of the grain boundary without a disordered layer. The tilt angle between the two grains was measured to be 36.8°. Its magnetic vs magnetoresistance curves measured at 2 K in Fig. 1(d-e) show that there are two peaks, which are very different from the SRO film without any grain boundaries[17]. Such transport behavior is characteristic of spin-valve magnetoresistance[12].

**Atomic-scale imaging of the GB.** To reveal the origin of spin-valve magnetoresistance at the grain boundary, we analyzed atomically resolved HAADF-STEM images to determine the atomic bonding of the SRO boundary shown in Fig. 2(a). Since HAADF shows Z contrast (Z is atomic number), the brighter spots in the image correspond to Ru columns, whereas the less bright spots are Sr columns. The (310) planes with different atomic termination layers appear at the boundary. The repeated units are marked by the white polygon. Fig. 2(b) is the atomistic mode that shows the cationic arrangements of the grain boundary core. Two Sr columns and one Ru column exist at the center of the polygon, which may be ascribed to the smaller radius of the Sr cation compared with that of the Ru cation[18]. To further confirm the atomic arrangement of the grain boundary, the atomically resolved EDS maps of Sr and Ru were recorded to verify the presence of any possible localized structural reconstruction, which commonly exist in the grain boundaries of complex oxides[19]. The net count maps of Sr (Fig. 2(c)), Ru (Fig. 2(d)), and the intermix of cations (Fig. 2(e)) are in excellent agreement with the structure model from the HAADF image, ruling out the existence of structural reconstruction in the grain boundary core.

However, the HAADF and EDS images in Fig. 2(a-e) show only cationic columns (Sr and Ru), and the scattering of O columns is too weak to be visible at such high collection angles. To determine the atomic arrangements of O at the boundary, an atomically resolved iDPC image was acquired, from which both cations and oxygen are visible[20]. Fig. 2(f) shows a typical iDPC-STEM image of the SRO grain boundary, with all the atomic columns of Sr, Ru and O being visible. Based on the contrast analysis of the iDPC image, the smallest (the weakest) dots in Fig. 2(f) represent the oxygen columns. The arrangements of the cations are consistent with those from the HAADF image in Fig. 2(a), and the positions of all oxygen columns in the grain boundary are identified as illustrated by the schematic in Fig. 2(g). Interestingly, the asymmetric grain boundary structure of SRO is different from the previously reported STO grain boundary[21]. The first principles calculations in Table S1 indicate that the asymmetric structure of the SRO grain boundary has a lower free energy than the symmetric ones. Therefore, the formation of the asymmetric SRO grain boundary is more favorable during growth.

**Distinct magnetic property at the boundary.** DFT calculations were also carried out to reveal the magnetic properties of the grain boundary. The details of the calculations are included in the methods section. The calculated Ru and O moments in $SrRuO_3$ bulk are 1.525 $\mu_B$ and 0.166 $\mu_B$, respectively, which are consistent with previous calculations and experiments[10]. Moreover, the projected density of states (DOS) in Fig. 3(b) and the band structure in Fig. S1 of bulk SrRuO$_3$ are also consistent with previous calculations[10]. The initial structure of the grain boundary is built based on the experimental data, and the relaxed grain boundary structure in Fig. 3(a) is in good agreement with the STEM images in Fig. 2(a) and Fig. 2(f). Ru and Sr are symmetric on the two sides of the boundary, which are defined as region A (above the grain boundary) and region B (below the grain boundary), respectively, as shown in Fig. 3(a). The DOS distribution of region B is similar to that of the bulk. However, for region A, the spin-up and spin-down DOS for Ru(O) become symmetrical, which is significantly different from the asymmetrical distribution in the bulk. Thus, the total magnetic moments at the grain boundary are significantly reduced compared to those in the bulk. The average magnetic moments for both region A and region B are calculated in Table S2. The magnetic moment is 0.134 $\mu_B$ for Ru and 0.012 $\mu_B$ for O in region A, while in region B, the magnetic moment is 1.166 $\mu_B$ for Ru and 0.105 $\mu_B$ for O. Moreover, along with the different magnetic moments on the two sides of the grain boundary, the spatial spin polarization (SSP) distributions on the two sides are distinct, as shown in Fig. 3(c). The positive spin polarization in region B is wider than that in region A.

To clarify the origin of the reduced and asymmetric distribution of moments and spin polarization in the SRO grain boundary, the energy band in Fig. S1, charge, orbital and octahedron distortion are analyzed in Fig. 4 and Fig. S2. The band structure of the SRO grain boundary shows the boundary region remains conductive. In Fig. 4(a), the spatial distribution of the charge density between region A and region B has no distinct difference. However, in Fig. 4(b), the spin-resolved charge differences in the two regions are completely different, which is consistent with the distinct moments on the two sides. The calculations show that the magnetic moment of Ru-6 is the smallest and that of Ru-1 is the largest. All five Ru $d$ orbital electrons ($d_{xy}$, $d_{yz}$, $d_{z^2}$, $d_{xz}$, and $d_{x^2-y^2}$) were calculated, as shown in Fig. 4(c), together with the bulk Ru. Orbital-projected DOS of Ru in region A and region B varied from the bulk case. The octahedral configurations shown in Fig. S2 and the length of the O octahedron edges listed in Table S3 suggest that the different O octahedral distortion results in Ru $d$ orbital reconstruction, leading to the distinct magnetic moments and spin polarization between the two sides of the SRO grain boundary.

The change in the magnetic property at the boundary certainly influences the transport properties of SRO containing the defects and interfacial properties of magnetoelectric heterostructure devices. The FM/NM/FM (FM = ferromagnetic, NM = nonmagnetic) sandwich structures formed across the grain boundary should have different transport properties than the pure FM phase. It was reported that TMR can be detected in the $La_{2/3}Sr_{1/3}MnO_3$ bicrystal[13] but not in the SRO bicrystal[12,13]. The TMR phenomenon existing in $La_{2/3}Sr_{1/3}MnO_3$ bicrystals is likely due to the magnetically disordered NM insulating layer formed at the grain boundary region serving as a tunneling barrier[1,22,23]. For SRO bicrystals, although an NM layer forms at the grain boundary, the metallic nature (as evidenced by the energy band in Fig. S1) instead of the insulating nature may lead to the large magnetoresistance[12,13]. Therefore, the measured transport property can be explained by the formation of the FM/NM/FM sandwich structure at the grain boundary, which is in excellent agreement with the theoretical calculations based on the atomic structure of the SRO grain boundary.

**DISCUSSION**

Previously, the grain boundaries in ferromagnetic $La_{0.7}Ca_{0.3}MnO_3$ and $La_{2/3}Sr_{1/3}MnO_3$ films were reported to significantly influence the magnetoresistance[1,2], likely due to the local transition from ferromagnetic to paramagnetic at the grain boundaries[3], while the dislocations in antiferromagnetic NiO were found to be ferromagnetic[24]. In this study, the grain boundary of ferromagnetic SRO becomes almost nonmagnetic. Therefore, it seems that the broken translation symmetry at the structural defects in these materials is usually accompanied by a change in the magnetic order. Since the structural defects can be zero-dimensional (e.g., point defect), one-dimensional (e.g., dislocation), and two-dimensional (e.g., grain boundary), it provides us strategies to design novel devices with low-dimensional magnetic order via proper defect engineering.

On the other hand, ferromagnetic SRO is widely used as an electrode for thin films such as ferroelectrics[25]. Considering that magnetoelectric heterostructures consisting of ferromagnetic and ferroelectric elements have recently aroused great interest due to their promising applications[26-28], the presence of NM grain boundaries in the FM SRO layers is expected to significantly change the interfacial magnetoelectric coupling.

**CONCLUSION**

In conclusion, we studied the atomic structure and magnetic and transport properties of the SRO Σ5(310) grain boundary. By using advanced atomically resolved iDPC images and EDS mapping, we were able to identify the atomic arrangements (including oxygen) at the grain boundary. We find that the structure of the grain boundary is asymmetric, which is very different from the common assumption based on the knowledge from STO. The DFT calculations show that the magnetic moments at the grain boundary are reduced, which originates from the distortion of the Ru-O octahedron-induced Ru $d$ orbital reconstruction. These results can well explain the observed transport properties, i.e., the spin-valve magnetoresistance at the grain boundary. This finding of the broken-translation-symmetry-induced change of the magnetic order at the grain boundary sheds light on the design of nanometer-scale devices with novel electronic/magnetic functions.

**METHODS**

**Thin film growth**. $SrRuO_3$ (SRO) thin films were deposited on $SrTiO_3$ (STO) bicrystal substrates with a tilt angle of 36.8° by the pulsed laser deposition technique using a KrF=248 nm excimer laser with a flux of approximately 5 J/cm$_2$ and a pulse repetition rate of 5 Hz. The $SrTiO_3$ bicrystal was purchased from Hefei Ke Jing Materials Technology Co., LTD. Before film deposition, the substrate temperature was raised to 700 °C with an oxygen pressure of 20 Pa. The deposition rate was set to approximately 0.5 nm min−1. Then, the films were cooled to room temperature. The X-ray diffraction pattern confirmed that the SRO thin film was grown on the STO substrate with a [001] epitaxial relationship.

**Magnetic and transport measurements.** The M-H curve is conducted by SQUID-VSM at a temperature of 2 K with an applied magnetic field from -5 T to 5 T. The Hall bar with a size of 5 μm was fabricated by electron beam lithography followed by Ar ion milling. The magnetoresistance is characterized by PPMS at 2 K with a magnetic field from -5 T to 5 T perpendicular to the film plane and a current of 100 μA.

**TEM sample preparation.** Thin foils for scanning transmission electron microscopy (STEM) observations were prepared by a conventional method that includes mechanical polishing of the sample back and then ion-beam milling. The ion-beam milling was carried out using argon ion milling (Leica EM RES102) with an acceleration voltage of 5 kV until a hole was made. Finally, low-voltage (0.8 kV) milling was carried out to reduce the irradiation-damaged layers.

**STEM characterization.** High-angle annular dark field (HADDF) images were recorded at 300 kV using an aberration-corrected FEI Titan Cube Themis G2 with a spatial resolution of approximately 60 pm. The convergence angle for imaging is 30 mrad, and the collection semiangle range is from 48 to 200 mrad. During imaging, low electron doses were applied using a small beam current (~50 pA) and a short scanning time. The energy dispersive X-ray spectroscopy (EDS) experiments were carried out at 300 kV by the Super EDS detectors.

A direct phase imaging technique, called integrated differential phase contrast (iDPC), was used, in which both heavy and light elements, including oxygen (nitrogen, carbon, etc.), are clearly visible[20]. In practice, iDPC-STEM is performed using a 4 quadrant (4Q) segmented detector, which enables an elegant solution for the thin sample transmission function phase problem because it is a very good approximation of an ideal center of mass (COM) or "first moment" detector. The iDPC image was recorded at 300 kV with a camera length of 350 mm and a DF4 detector (collection semiangle: 5 to 27 mrad).

**DFT simulations.** The first-principle calculations are implemented with the Vienna Ab initio simulation package with density functional theory (DFT)[29]. The Perdew-Burke-Ernzerhof spin-polarized generalized gradient approximation and projector augment wave pseudopotentials are applied. The energy cutoff for the plane wave basis set is 500 eV[10]. The convergence criteria for the energy and atomic forces are $10^{-5}$ eV and 0.01 eV/Å, respectively. The Brillouin Zone is sampled with Γ-centered 9×9×9 and 9×3×2 $k$ point meshes for the $SrRuO_3$ bulk and grain boundary models, respectively. The grain boundary model is built by the $SrRuO_3$ (310) plane. The spatial spin polarization[30] is defined as:

$$P(r,z,\varepsilon) = \frac{n_s^\uparrow(r,z,\varepsilon) - n_s^\downarrow(r,z,\varepsilon)}{n_s^\uparrow(r,z,\varepsilon) + n_s^\downarrow(r,z,\varepsilon)} \qquad (1)$$

where $n_s^{\uparrow(\downarrow)}(r,z,\varepsilon)$ is the spin-up (down) charge density in real space with an energy interval of [$\varepsilon$, $E_F$]. The spin-resolved charge difference is calculated by subtracting the spin-down charge density from the spin-up charge density.

**Acknowledgements**

The authors acknowledge Electron Microscopy Laboratory in Peking University for the use of Cs corrected electron microscope.

**Funding**

This work was supported by the National Key R&D Program of China (2016YFA0300804), National Equipment Program of China (ZDYZ2015-1), National Natural Science Foundation of China (51672007, 51502007, 11327902, 11474337, 21773303 and 51421002), the National Program for Thousand Young Talents of China and "2011 Program" Peking-Tsinghua-IOP Collaborative Innovation Center of Quantum Matter.

Conflict of interest statement. None declared.

**Figures**

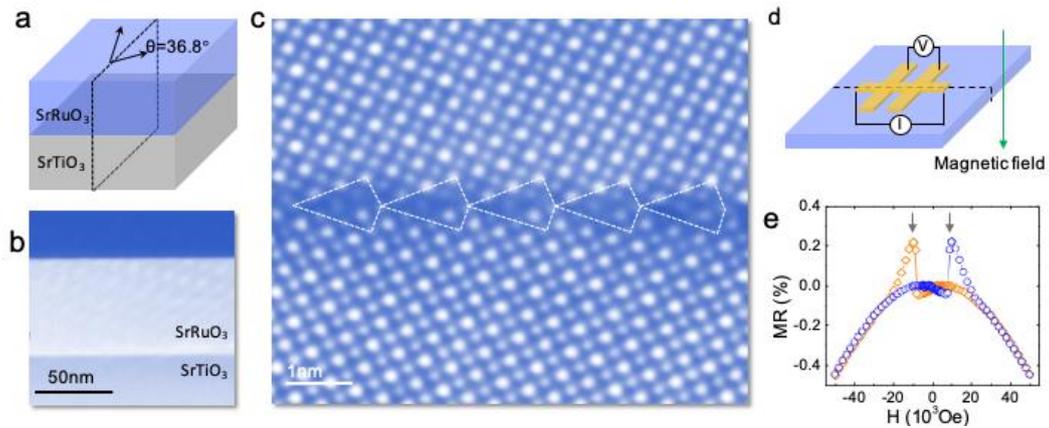

**Figure 1.** SrRuO3 (SRO) bicrystal fabrication and characterization. (a) A schematic diagram showing the SRO bicrystal film on the SrTiO3 bicrystal substrate. (b) A HAADF-STEM image of a cross-sectional sample showing a thin film thickness of approximately 50 nm. (c) A HAADF-STEM image of the planar view sample showing a high-quality 36.8° grain boundary without a disordered layer. (d) Schematic showing the device for the transport measurement. (e) Resistance is plotted as a function of magnetic field at 2 K. The two labeled peaks characterize the spin-valve magnetoresistance.

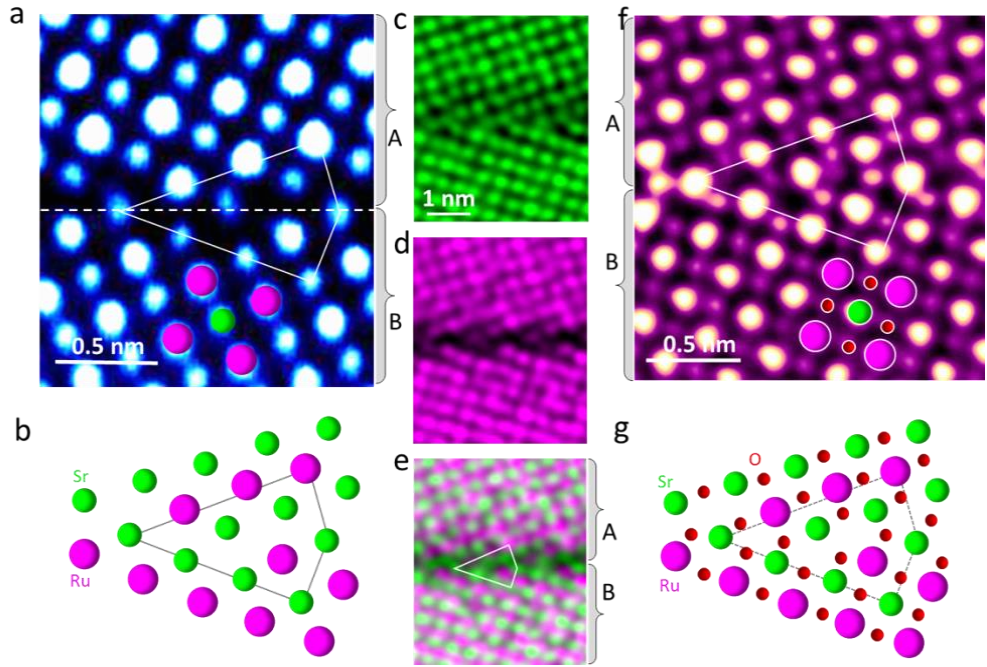

**Figure 2.** Atomic arrangements of the grain boundary. (a) An atomically resolved HAADF-STEM image of the grain boundary along the [001] direction. The above part is defined as region A, and the below part is defined as region B hereinafter. (b) Schematic showing the cationic arrangements along the [001] direction based on the HAADF image, Sr (green) and Ru (pink). (c-e) Net count maps of (c) Sr (green), (d) Ru (pink), and (**e**) the intermix of Sr and Ru showing no structural reconstruction at the grain boundary. (f) An atomically resolved iDPC STEM image showing the anionic and cationic configuration. (g) Schematic representation of the atomic structure of the SRO grain boundary, Sr (green), Ru (pink) and O (red).

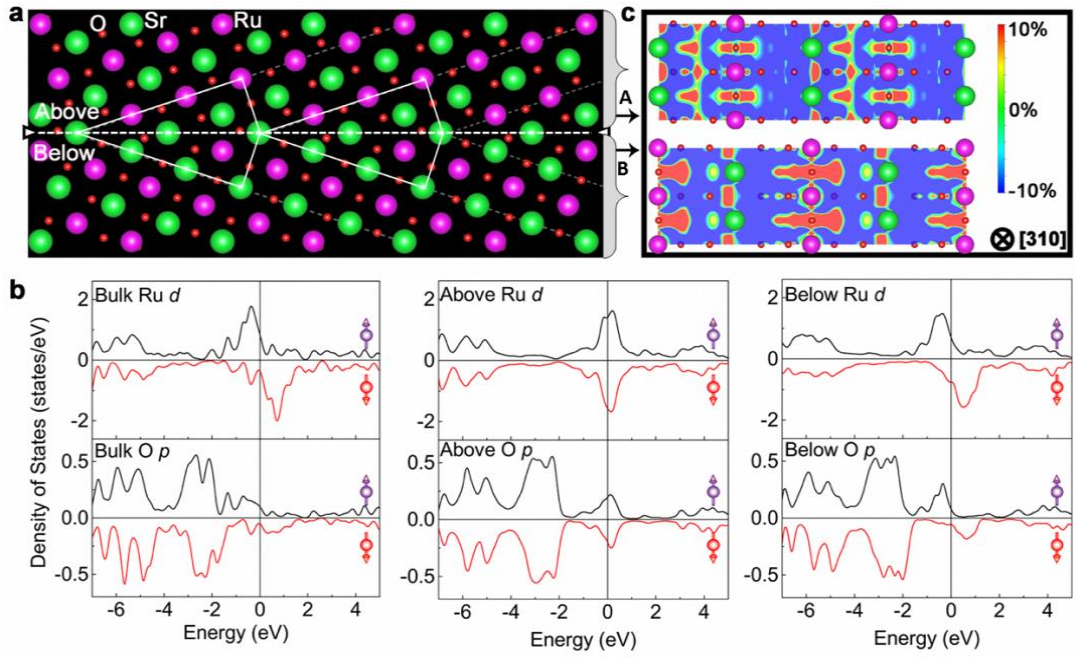

**Figure 3.** DFT calculation of the SRO grain boundary. (a) Relaxed structure of the SRO grain boundary. (b) Projected density of states (DOS) for the SRO bulk and grain boundary. The Fermi level ($E_F$) is indicated by the vertical lines that are set to zero. (c) The spatial spin polarization in the above and below layers of the SRO grain boundary model in the (310) plane with the energy interval of [$E_F$-0.2 eV, $E_F$].

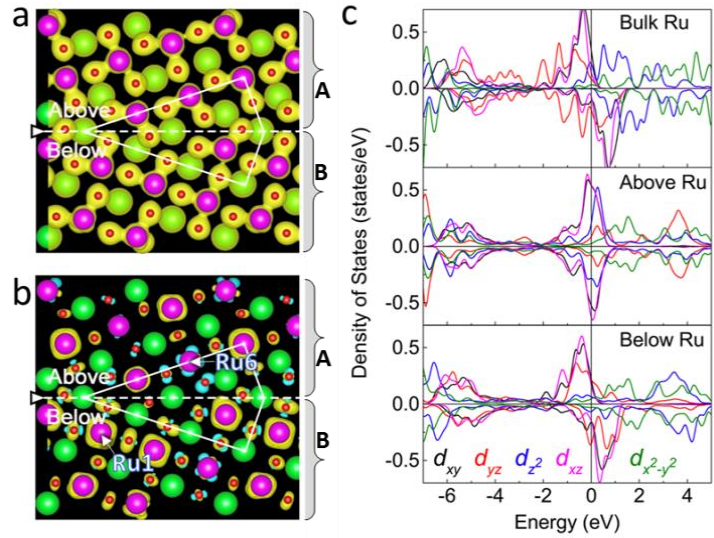

**Figure 4.** Charge and spin distribution at the SRO grain boundary. (a) Spatial distribution of charge density in the grain boundary structure (isosurface value 0.08 e/Å$_3$). (b) Spatial distribution of the spin-resolved charge difference in the grain boundary structure (isosurface value 0.002 e/Å$_3$). The yellow color indicates the large spin-up charge density, and blue represents the large spin-down charge density. (c) Orbital-projected DOS of Ru in the SRO bulk and grain boundary model.

Supplementary Information for

# Atomic Origin of Spin-Valve Magnetoresistance at the SrRuO$_3$ Grain Boundary

Xujing Li[1,2,‡], Li Yin[3,‡], Zhengxun Lai[3], Mei Wu[1,4], Yu Sheng[5], Lei Zhang[2], Yuanwei Sun[1,4], Shulin Chen[1], Xiaomei Li[2], Jingmin Zhang[1], Yuehui Li[1,4], Kaihui Liu[6,7], Kaiyou Wang[5,8], Dapeng Yu[1,6,9], Xuedong Bai[2,7],*, Wenbo Mi[3],*, Peng Gao[1,4,7],*

[1]Electron Microscopy Laboratory, School of Physics, Peking University, Beijing, 100871, China

[2]Beijing National Laboratory for Condensed Matter Physics and Institute of Physics, Chinese Academy of Sciences, Beijing 100190, China

[3]Tianjin Key Laboratory of Low Dimensional Materials Physics and Preparation Technology, School of Science, Tianjin University, Tianjin 300354, China

[4]International Center for Quantum Materials, School of Physics, Peking University, Beijing, 100871, China

[5]State Key Laboratory of Superlattices and Microstructures, Institute of Semiconductors, Chinese Academy of Sciences, Beijing 100083, China

[6] State Key Laboratory for Artificial Microstructure & Mesoscopic Physics, School of Physics, Peking University, Beijing 100871, China

[7]Collaborative Innovation Centre of Quantum Matter, Beijing 100871, China

[8] College of Materials science and Opto-Electronic Technology, University of Chinese Academy of Sciences, Beijing 100049, P. R. China

[9]Shenzhen Institute for Quantum Science and Engineering (SIQSE), and Department of Physics, Southern University of Science and Technology (SUSTech), Shenzhen 518055, P.R.China.

‡These authors contributed equally to this work.
E-mails: p-gao@pku.edu.cn; miwenbo@tju.edu.cn; xdbai@iphy.ac.cn

## S1. Simulation of symmetric and asymmetric structure

In Table S1, a is the model obtained in our experiment, b is the model of symmetric O ions but asymmetric cations on both sides of grain boundary, and c is the model of symmetric O ions and cations on both sides of grain boundary, which is built based on the atomic structure of the same type of SrTiO$_3$ grain boundaries reported in literature[1]. The Table S1 suggests that the asymmetric model 'a' has the lowest free energy and is more stable than the other two models.

| | a | b | c |
|---|---|---|---|
| model | 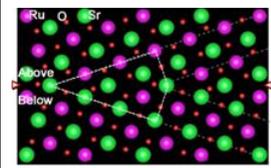 | 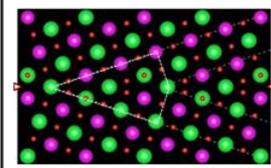 | 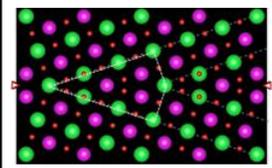 |
| Energy(J/m$^2$) | -131.804 | -125.883 | -126.016 |

**Table** S1. DFT calculations of free energy for grain boundaries with different structures. Structures for (a) asymmetric, (b) symmetric O but asymmetric cations, (c) symmetric O and symmetric cations of SRO grain boundaries and the corresponding free energy.

## S2. Band structure

The band structure of SRO grain boundary is intensively distributed compared with the bulk, but both exhibiting conducting properties.

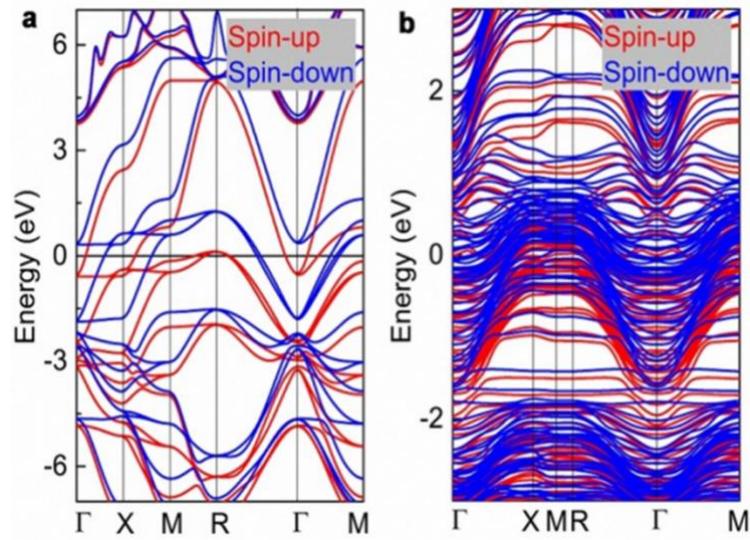

**Figure** S1. DFT calculations of band structure. Band structures of (a) SRO bulk and (b) grain boundary.

**S3. The magnetic moment**

|  | $M_{Ru}$ ($\mu_B$) | $M_O$ ($\mu_B$) |
|---|---|---|
| Bulk | 1.525 | 0.166 |
| Above | 0.134 | 0.012 |
| Below | 1.166 | 0.105 |

**Table** S2. DFT calculations of magnetic moments. Averaged magnetic moment ($\mu_B$) of Ru and O atoms in SRO bulk and grain boundary.

## S4. Length of O octahedron edges

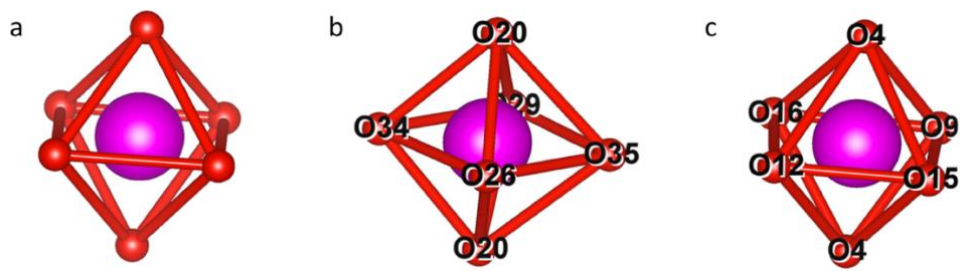

**Figure** S2. DFT calculations of RuO$_6$ octahedron near the grain boundary. The RuO$_6$ octahedral configurations of (a) bulk, (b) Ru-6, and (c) Ru-1.

| Model | Atoms | Distance | Experimental | Model | Atoms | Distance | Experimental |
|---|---|---|---|---|---|---|---|
| Ru1 (Below) | O4-O16 | 2.778 | | Ru-6 (Above) | O20-O34 | 2.777 | |
| | O4-O12 | 2.906 | | | O20-O26 | 2.738 | |
| | O4-O15 | 2.878 | | | O20-O35 | 2.925 | |
| | O4-O9 | 2.878 | | | O20-O29 | 2.861 | |
| | O16-O12 | 2.844 | 2.839 | | O34-O26 | 2.831 | 2.792 |
| | O12-O15 | 2.698 | 2.703 | | O26-O35 | 2.716 | 2.684 |
| | O15-O9 | 2.812 | 2.804 | | O35-O29 | 2.799 | 2.764 |
| | O9-O16 | 2.885 | 2.891 | | O29-O34 | 2.888 | 2.834 |
| Bulk | - | 2.836 | | | | | |

**Table** S3. The calculated atom distances. The distances (Å) of O octahedron for Ru1 and Ru 6 compared with the bulk.

**Supplementary references**